\documentclass[aps,pre, 
longbibliography,
reprint,   
twocolumn, 
nofootinbib,
floatfix]{revtex4-2}

\usepackage[T1]{fontenc}
\usepackage[utf8]{inputenc}
\usepackage{amsmath}
\usepackage{amssymb}
\usepackage{graphicx}
\usepackage{dcolumn}
\usepackage{caption}
\usepackage{xcolor}
\usepackage[labelformat=parens]{subcaption}
\DeclareCaptionSubType*[arabic]{figure}

\usepackage{listings}
\definecolor{codegreen}{rgb}{0,0.6,0}
\definecolor{codegray}{rgb}{0.5,0.5,0.5}
\definecolor{codepurple}{rgb}{0.58,0,0.82}
\definecolor{backcolour}{rgb}{0.95,0.95,0.92}
\lstdefinestyle{mystyle}{
    backgroundcolor=\color{backcolour},   
    commentstyle=\color{codegreen},
    keywordstyle=\color{magenta},
    numberstyle=\tiny\color{codegray},
    stringstyle=\color{codepurple},
    basicstyle=\ttfamily\footnotesize,
    breakatwhitespace=false,         
    breaklines=true,                 
    captionpos=t,                    
    keepspaces=true,                 
    numbers=left,                    
    numbersep=5pt,                  
    showspaces=false,                
    showstringspaces=false,
    showtabs=false,                  
    tabsize=2
}
\usepackage[colorlinks=true,hyperfootnotes=true,breaklinks=true,allcolors=teal]{hyperref}
\usepackage{cleveref}

\begin{document}
\title{Lower limit of percolation threshold on a square lattice with complex neighborhoods}
\author{Antoni Ciepłucha}
\thanks{\href{https://orcid.org/0009-0006-4018-7027}{0009-0006-4018-7027}}
\author{Marcin Utnicki}
\author{Maciej Wołoszyn}
\thanks{\href{https://orcid.org/0000-0001-9896-1018}{0000-0001-9896-1018}}
\author{Krzysztof Malarz}
\thanks{\href{https://orcid.org/0000-0001-9980-0363}{0000-0001-9980-0363}}
\email{malarz@agh.edu.pl}
\affiliation{\href{https://ror.org/00bas1c41}{AGH University}, Faculty of Physics and Applied Computer Science, al.~Mickiewicza~30, 30-059 Krak\'ow, Poland}
\date{\today}

\begin{abstract}
In this paper, the 60-year-old concept of long-range interaction in percolation problems introduced by Dalton, Domb, and Sykes, is reconsidered. 
With Monte Carlo simulation---based on Newman--Ziff algorithm and finite-size scaling hypothesis---we estimate 64 percolation thresholds for random site percolation problem on a square lattice with neighborhoods that contain sites from the 7th coordination zone.
The percolation thresholds obtained range from $0.27013$ (for the neighborhood that contains only sites from the 7th coordination zone) to $0.11535$ (for the neighborhood that contains all sites from the 1st to the 7th coordination zone).
Similarly to neighborhoods with smaller ranges, the power-law dependence of the percolation threshold on the effective coordination number with exponent close to $-1/2$ is observed.
Finally, we empirically determine the limit of the percolation threshold on square lattices with complex neighborhoods.
This limit scales with the inverse square of the mean radius of the neighborhood.
The boundary of this limit is touched for threshold values associated with extended (compact) neighborhoods.
\end{abstract}

\maketitle

\section{Introduction}

The percolation \cite{bookDS,bookBB,bookMS,bookHK} (see References~\cite{Saberi2015,Li_2021} for recent reviews) is one of the core topics in statistical physics providing the possibility to look at the critical phenomena occurring at phase transition solely on the geometrical basis, without samples heating, cooling, inserting in a magnetic field, etc.
Some problems with respect to percolation can be treated analytically \cite{Sykes_1964,Ziff_2006,PhysRevE.73.016107,Jacobsen_2014,PhysRevE.105.044108,Akhunzhanov_2022}, but most studies are computational.

Originating from works by \citeauthor{Broadbent1957} \cite{Broadbent1957,Hammersley1957} devoted to rheology (and still applied there \cite{ISI:000496837300028,Bolandtaba2011,Mun2014,ISI:000524118200031}), it quickly found plenty of applications in physics, including magnetic \cite{PhysRevB.97.165121,ISI:000400959000004,PhysRevB.94.054407,Yiu,Grady_2023} and electric \cite{ISI:000419615800018,ISI:000462936100013,ISI:000514848600043} properties of solids or nano-engineering \cite{Xu_2014}.

However, the application of percolation theory is not restricted to physics alone.
Such examples (mainly two-dimensional systems \cite{Sykes1976a,Sykes1976b,Sykes1976c,Sykes1976d,Gaunt1976}) can also be found in
epidemiology \cite{Meyers_2007,Lee_2021,2101.00550},
forest fires \cite{Kaczanowska2002,Guisoni2011,Simeoni2011,Camelo-Neto2011,Abades2014},
agriculture \cite{ISI:000518460000003,Rosales_Herrera_2021,PhysRevE.109.014304},
urbanization \cite{ISI:000523958600016,Ng_2024},
materials chemistry \cite{Alguero_2020},
sociology \cite{ISI:000168785900005},
psychology \cite{2206.14226},
information transfer \cite{Cirigliano_2023},
finances \cite{Bartolucci2020},
or dentistry \cite{Beddoe_2023}.

On the other hand, some effort was put into studying percolation on cubic ($d=3$ \cite{PhysRevE.67.036101,PhysRev.133.A310,Sur_1976,Gaunt_1983,Lorenz_1998,Kurzawski2012}) and hyper-cubic lattices, also for non-physical dimensions ($d=4$ \cite{1803.09504,PhysRevE.67.036101,Zhao_2022,PhysRevE.64.026115}, $d=5$ \cite{2308.15719,PhysRevE.67.036101,PhysRevE.64.026115} and higher $d\ge 6$ \cite{Van_der_Marck_1998,PhysRevE.67.036101,Koza_2016}).
Simultaneously, the complex \cite{Li_2021,Cirigliano_2023,PhysRevE.79.021118}, distorted \cite{PhysRevE.99.012117,PhysRevE.106.034109,2306.05513} and fractal \cite{fractalfract7030231} networks were studied.

Recently, the 60-year-old concept of long-range interaction \cite{Dalton_1964,Domb1966} has sparked renewed interest.
The neighborhood that contains sites that are not nearest-neighbors on assumed lattice are called extended neighborhoods (when they are compact) or complex (when they are non-compact, that is, they contain `wholes')---to keep the nomenclature from Reference~\cite{2310.20668}.
Many papers were devoted to studies of the percolation thresholds on the extended \cite{PhysRevE.103.022126,PhysRevE.105.024105,Zhao_2022,Xun_2022,Cirigliano_2023,2308.15719} and the complex \cite{Galam2005a,Galam2005b,Majewski2007,Kurzawski2012,Malarz2015,1803.09504,2006.15621,2102.10066,2204.12593,2303.10423,2310.20668} neighborhoods.

The particular neighborhoods names---now keeping the convention proposed in Reference~\cite{2010.02895}---are combination of alphanumeric strings. The first two letters identify the underlying lattice (for example: \textsc{sq} for square, \textsc{tr} for triangular, \textsc{hc} for honeycomb, and \textsc{sc} for simple cubic lattices) then they are accompanied with numerical string indicating coordination zones which constitute the neighborhood.
In this convention, the von Neumann neighborhood on the square lattice (with only the nearest neighbors) is called \textsc{sq-1}, while Moore's neighborhood on square lattice (containing sites from the first and the second coordination zones) is called \textsc{sq-1,2}.

In this paper we are closer to the theoretical studies of percolation phenomena than to its application.
That is, our studies focus on the influence of long-range interactions on the percolation threshold $p_c$.
The percolation threshold is the equivalent of the critical point in the phase-transition phenomena.
For the random site percolation problem, we deal with the nodes of the lattice that are occupied (with probability $p$) or empty (with probability $1-p$).
Occupied sites, in assumed neighborhood, are considered to form a cluster.
Depending on the sites occupation probability $p$ such cluster may span (or not) the system edges.
The percolation threshold $p_c$ is such an occupation probability $p$, that for $p<p_c$ a spanning cluster is absent (and the system behaves as an insulator), while for $p>p_c$ a spanning cluster is present (and the system behaves as a conductor).
Thus, $p_c$ separates two phases: isolating and conducting, and at $p=p_c$ the (second-order) phase transition takes place.

\begin{figure*}[htbp]
\begin{subfigure}[b]{0.090\textwidth}
\caption{\label{fig:sq-7}}
\includegraphics[width=\textwidth]{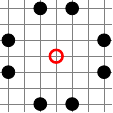} 
\end{subfigure}
\hfill 
\begin{subfigure}[b]{0.090\textwidth}
\caption{\label{fig:sq-17}}
\includegraphics[width=\textwidth]{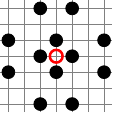} 
\end{subfigure}
\hfill 
\begin{subfigure}[b]{0.090\textwidth}
\caption{\label{fig:sq-27}}
\includegraphics[width=\textwidth]{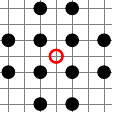} 
\end{subfigure}
\hfill 
\begin{subfigure}[b]{0.090\textwidth}
\caption{\label{fig:sq-37}}
\includegraphics[width=\textwidth]{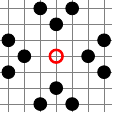} 
\end{subfigure}
\hfill 
\begin{subfigure}[b]{0.090\textwidth}
\caption{\label{fig:sq-47}}
\includegraphics[width=\textwidth]{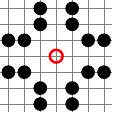} 
\end{subfigure}
\hfill 
\begin{subfigure}[b]{0.090\textwidth}
\caption{\label{fig:sq-57}}
\includegraphics[width=\textwidth]{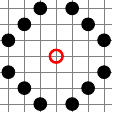} 
\end{subfigure}
\hfill 
\begin{subfigure}[b]{0.090\textwidth}
\caption{\label{fig:sq-67}}
\includegraphics[width=\textwidth]{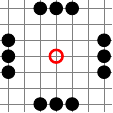} 
\end{subfigure}
\hfill 
\begin{subfigure}[b]{0.090\textwidth}
\caption{\label{fig:sq-127}}
\includegraphics[width=\textwidth]{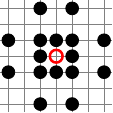} 
\end{subfigure}\\
\begin{subfigure}[b]{0.090\textwidth}
\caption{\label{fig:sq-137}}
\includegraphics[width=\textwidth]{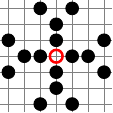} 
\end{subfigure}
\hfill 
\begin{subfigure}[b]{0.090\textwidth}
\caption{\label{fig:sq-147}}
\includegraphics[width=\textwidth]{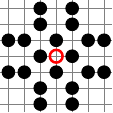} 
\end{subfigure}
\hfill 
\begin{subfigure}[b]{0.090\textwidth}
\caption{\label{fig:sq-157}}
\includegraphics[width=\textwidth]{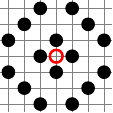} 
\end{subfigure}
\hfill 
\begin{subfigure}[b]{0.090\textwidth}
\caption{\label{fig:sq-167}}
\includegraphics[width=\textwidth]{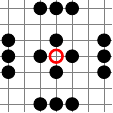} 
\end{subfigure}
\hfill 
\begin{subfigure}[b]{0.090\textwidth}
\caption{\label{fig:sq-237}}
\includegraphics[width=\textwidth]{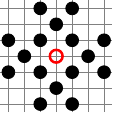} 
\end{subfigure}
\hfill 
\begin{subfigure}[b]{0.090\textwidth}
\caption{\label{fig:sq-247}}
\includegraphics[width=\textwidth]{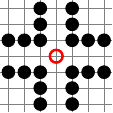} 
\end{subfigure}
\hfill 
\begin{subfigure}[b]{0.090\textwidth}
\caption{\label{fig:sq-257}}
\includegraphics[width=\textwidth]{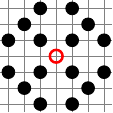} 
\end{subfigure}
\hfill 
\begin{subfigure}[b]{0.090\textwidth}
\caption{\label{fig:sq-267}}
\includegraphics[width=\textwidth]{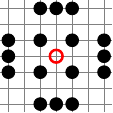} 
\end{subfigure}\\
\begin{subfigure}[b]{0.090\textwidth}
\caption{\label{fig:sq-347}}
\includegraphics[width=\textwidth]{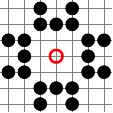} 
\end{subfigure}
\hfill 
\begin{subfigure}[b]{0.090\textwidth}
\caption{\label{fig:sq-357}}
\includegraphics[width=\textwidth]{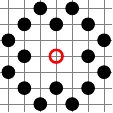} 
\end{subfigure}
\hfill 
\begin{subfigure}[b]{0.090\textwidth}
\caption{\label{fig:sq-367}}
\includegraphics[width=\textwidth]{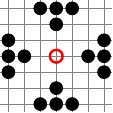} 
\end{subfigure}
\hfill 
\begin{subfigure}[b]{0.090\textwidth}
\caption{\label{fig:sq-457}}
\includegraphics[width=\textwidth]{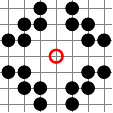} 
\end{subfigure}
\hfill 
\begin{subfigure}[b]{0.090\textwidth}
\caption{\label{fig:sq-467}}
\includegraphics[width=\textwidth]{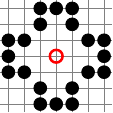} 
\end{subfigure}
\hfill 
\begin{subfigure}[b]{0.090\textwidth}
\caption{\label{fig:sq-567}}
\includegraphics[width=\textwidth]{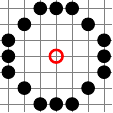} 
\end{subfigure}
\hfill 
\begin{subfigure}[b]{0.090\textwidth}
\caption{\label{fig:sq-1237}}
\includegraphics[width=\textwidth]{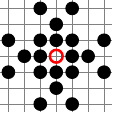} 
\end{subfigure}
\hfill 
\begin{subfigure}[b]{0.090\textwidth}
\caption{\label{fig:sq-1247}}
\includegraphics[width=\textwidth]{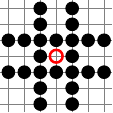} 
\end{subfigure}\\
\begin{subfigure}[b]{0.090\textwidth}
\caption{\label{fig:sq-1257}}
\includegraphics[width=\textwidth]{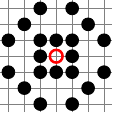} 
\end{subfigure}
\hfill 
\begin{subfigure}[b]{0.090\textwidth}
\caption{\label{fig:sq-1267}}
\includegraphics[width=\textwidth]{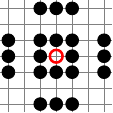} 
\end{subfigure}
\hfill 
\begin{subfigure}[b]{0.090\textwidth}
\caption{\label{fig:sq-1347}}
\includegraphics[width=\textwidth]{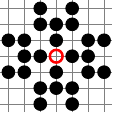} 
\end{subfigure}
\hfill 
\begin{subfigure}[b]{0.090\textwidth}
\caption{\label{fig:sq-1357}}
\includegraphics[width=\textwidth]{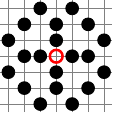} 
\end{subfigure}
\hfill 
\begin{subfigure}[b]{0.090\textwidth}
\caption{\label{fig:sq-1367}}
\includegraphics[width=\textwidth]{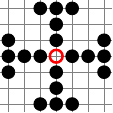} 
\end{subfigure}
\hfill 
\begin{subfigure}[b]{0.090\textwidth}
\caption{\label{fig:sq-1457}}
\includegraphics[width=\textwidth]{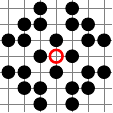} 
\end{subfigure}
\hfill 
\begin{subfigure}[b]{0.090\textwidth}
\caption{\label{fig:sq-1467}}
\includegraphics[width=\textwidth]{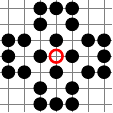} 
\end{subfigure}
\hfill 
\begin{subfigure}[b]{0.090\textwidth}
\caption{\label{fig:sq-1567}}
\includegraphics[width=\textwidth]{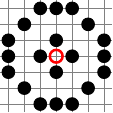} 
\end{subfigure}\\
\begin{subfigure}[b]{0.090\textwidth}
\caption{\label{fig:sq-2347}}
\includegraphics[width=\textwidth]{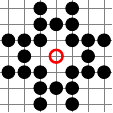} 
\end{subfigure}
\hfill 
\begin{subfigure}[b]{0.090\textwidth}
\caption{\label{fig:sq-2357}}
\includegraphics[width=\textwidth]{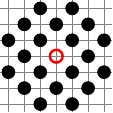} 
\end{subfigure}
\hfill 
\begin{subfigure}[b]{0.090\textwidth}
\caption{\label{fig:sq-2367}}
\includegraphics[width=\textwidth]{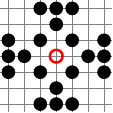} 
\end{subfigure}
\hfill 
\begin{subfigure}[b]{0.090\textwidth}
\caption{\label{fig:sq-2457}}
\includegraphics[width=\textwidth]{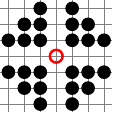} 
\end{subfigure}
\hfill 
\begin{subfigure}[b]{0.090\textwidth}
\caption{\label{fig:sq-2467}}
\includegraphics[width=\textwidth]{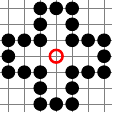} 
\end{subfigure}
\hfill 
\begin{subfigure}[b]{0.090\textwidth}
\caption{\label{fig:sq-2567}}
\includegraphics[width=\textwidth]{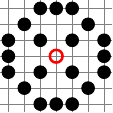} 
\end{subfigure}
\hfill 
\begin{subfigure}[b]{0.090\textwidth}
\caption{\label{fig:sq-3457}}
\includegraphics[width=\textwidth]{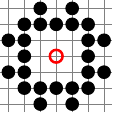} 
\end{subfigure}
\hfill 
\begin{subfigure}[b]{0.090\textwidth}
\caption{\label{fig:sq-3467}}
\includegraphics[width=\textwidth]{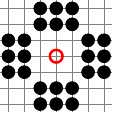} 
\end{subfigure}\\
\begin{subfigure}[b]{0.090\textwidth}
\caption{\label{fig:sq-3567}}
\includegraphics[width=\textwidth]{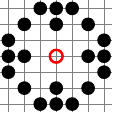} 
\end{subfigure}
\hfill 
\begin{subfigure}[b]{0.090\textwidth}
\caption{\label{fig:sq-4567}}
\includegraphics[width=\textwidth]{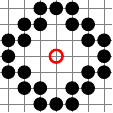} 
\end{subfigure}
\hfill 
\begin{subfigure}[b]{0.090\textwidth}
\caption{\label{fig:sq-12347}}
\includegraphics[width=\textwidth]{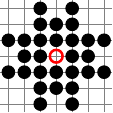} 
\end{subfigure}
\hfill 
\begin{subfigure}[b]{0.090\textwidth}
\caption{\label{fig:sq-12357}}
\includegraphics[width=\textwidth]{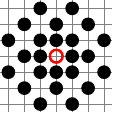} 
\end{subfigure}
\hfill 
\begin{subfigure}[b]{0.090\textwidth}
\caption{\label{fig:sq-12367}}
\includegraphics[width=\textwidth]{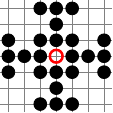} 
\end{subfigure}
\hfill 
\begin{subfigure}[b]{0.090\textwidth}
\caption{\label{fig:sq-12457}}
\includegraphics[width=\textwidth]{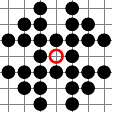} 
\end{subfigure}
\hfill 
\begin{subfigure}[b]{0.090\textwidth}
\caption{\label{fig:sq-12467}}
\includegraphics[width=\textwidth]{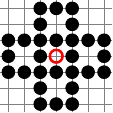} 
\end{subfigure}
\hfill 
\begin{subfigure}[b]{0.090\textwidth}
\caption{\label{fig:sq-12567}}
\includegraphics[width=\textwidth]{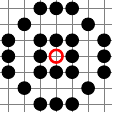} 
\end{subfigure}\\
\begin{subfigure}[b]{0.090\textwidth}
\caption{\label{fig:sq-13457}}
\includegraphics[width=\textwidth]{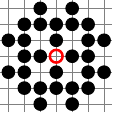} 
\end{subfigure}
\hfill 
\begin{subfigure}[b]{0.090\textwidth}
\caption{\label{fig:sq-13467}}
\includegraphics[width=\textwidth]{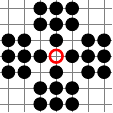} 
\end{subfigure}
\hfill 
\begin{subfigure}[b]{0.090\textwidth}
\caption{\label{fig:sq-13567}}
\includegraphics[width=\textwidth]{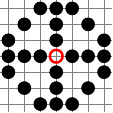} 
\end{subfigure}
\hfill 
\begin{subfigure}[b]{0.090\textwidth}
\caption{\label{fig:sq-14567}}
\includegraphics[width=\textwidth]{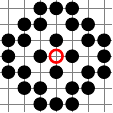} 
\end{subfigure}
\hfill 
\begin{subfigure}[b]{0.090\textwidth}
\caption{\label{fig:sq-23457}}
\includegraphics[width=\textwidth]{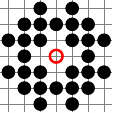} 
\end{subfigure}
\hfill 
\begin{subfigure}[b]{0.090\textwidth}
\caption{\label{fig:sq-23467}}
\includegraphics[width=\textwidth]{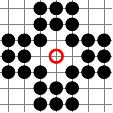} 
\end{subfigure}
\hfill 
\begin{subfigure}[b]{0.090\textwidth}
\caption{\label{fig:sq-23567}}
\includegraphics[width=\textwidth]{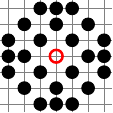} 
\end{subfigure}
\hfill 
\begin{subfigure}[b]{0.090\textwidth}
\caption{\label{fig:sq-24567}}
\includegraphics[width=\textwidth]{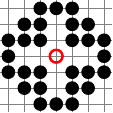} 
\end{subfigure}\\
\begin{subfigure}[b]{0.090\textwidth}
\caption{\label{fig:sq-34567}}
\includegraphics[width=\textwidth]{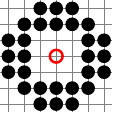} 
\end{subfigure}
\hfill 
\begin{subfigure}[b]{0.090\textwidth}
\caption{\label{fig:sq-123457}}
\includegraphics[width=\textwidth]{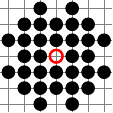} 
\end{subfigure}
\hfill 
\begin{subfigure}[b]{0.090\textwidth}
\caption{\label{fig:sq-123467}}
\includegraphics[width=\textwidth]{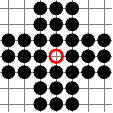} 
\end{subfigure}
\hfill 
\begin{subfigure}[b]{0.090\textwidth}
\caption{\label{fig:sq-123567}}
\includegraphics[width=\textwidth]{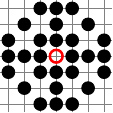} 
\end{subfigure}
\hfill 
\begin{subfigure}[b]{0.090\textwidth}
\caption{\label{fig:sq-124567}}
\includegraphics[width=\textwidth]{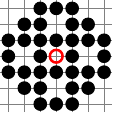} 
\end{subfigure}
\hfill 
\begin{subfigure}[b]{0.090\textwidth}
\caption{\label{fig:sq-134567}}
\includegraphics[width=\textwidth]{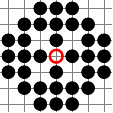} 
\end{subfigure}
\hfill 
\begin{subfigure}[b]{0.090\textwidth}
\caption{\label{fig:sq-234567}}
\includegraphics[width=\textwidth]{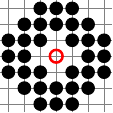} 
\end{subfigure}
\hfill 
\begin{subfigure}[b]{0.090\textwidth}
\caption{\label{fig:sq-1234567}}
\includegraphics[width=\textwidth]{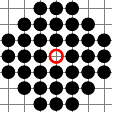} 
\end{subfigure}
\caption{\label{fig:neighborhoods}Neighborhoods on square lattice containing sites from the 7-th coordination zone:
(\subref{fig:sq-7}) \textsc{sq-7},
(\subref{fig:sq-17}) \textsc{sq-1,7},
(\subref{fig:sq-27}) \textsc{sq-2,7},
(\subref{fig:sq-37}) \textsc{sq-3,7},
(\subref{fig:sq-47}) \textsc{sq-4,7},
(\subref{fig:sq-57}) \textsc{sq-5,7},
(\subref{fig:sq-67}) \textsc{sq-6,7},
(\subref{fig:sq-127}) \textsc{sq-1,2,7},
(\subref{fig:sq-157}) \textsc{sq-1,5,7},
(\subref{fig:sq-167}) \textsc{sq-1,6,7},
(\subref{fig:sq-237}) \textsc{sq-2,3,7},
(\subref{fig:sq-247}) \textsc{sq-2,4,7},
(\subref{fig:sq-257}) \textsc{sq-2,5,7},
(\subref{fig:sq-267}) \textsc{sq-2,6,7},
(\subref{fig:sq-347}) \textsc{sq-3,4,7},
(\subref{fig:sq-357}) \textsc{sq-3,5,7},
(\subref{fig:sq-367}) \textsc{sq-3,6,7},
(\subref{fig:sq-457}) \textsc{sq-4,5,7},
(\subref{fig:sq-467}) \textsc{sq-4,6,7},
(\subref{fig:sq-567}) \textsc{sq-5,6,7},
(\subref{fig:sq-1237}) \textsc{sq-1,2,3,7},
(\subref{fig:sq-1247}) \textsc{sq-1,2,4,7},
(\subref{fig:sq-1257}) \textsc{sq-1,2,5,7},
(\subref{fig:sq-1267}) \textsc{sq-1,2,6,7},
(\subref{fig:sq-1347}) \textsc{sq-1,3,4,7},
(\subref{fig:sq-1357}) \textsc{sq-1,3,5,7},
(\subref{fig:sq-1367}) \textsc{sq-1,3,6,7},
(\subref{fig:sq-1457}) \textsc{sq-1,4,5,7},
(\subref{fig:sq-1467}) \textsc{sq-1,4,6,7},
(\subref{fig:sq-1567}) \textsc{sq-1,5,6,7},
(\subref{fig:sq-2347}) \textsc{sq-2,3,4,7},
(\subref{fig:sq-2357}) \textsc{sq-2,3,5,7},
(\subref{fig:sq-2367}) \textsc{sq-2,3,6,7},
(\subref{fig:sq-2457}) \textsc{sq-2,4,5,7},
(\subref{fig:sq-2467}) \textsc{sq-2,4,6,7},
(\subref{fig:sq-2567}) \textsc{sq-2,5,6,7},
(\subref{fig:sq-3457}) \textsc{sq-3,4,5,7},
(\subref{fig:sq-3467}) \textsc{sq-3,4,6,7},
(\subref{fig:sq-3567}) \textsc{sq-3,5,6,7},
(\subref{fig:sq-4567}) \textsc{sq-4,5,6,7},
(\subref{fig:sq-12347}) \textsc{sq-1,2,3,4,7},
(\subref{fig:sq-12357}) \textsc{sq-1,2,3,5,7},
(\subref{fig:sq-12367}) \textsc{sq-1,2,3,6,7},
(\subref{fig:sq-12457}) \textsc{sq-1,2,4,5,7},
(\subref{fig:sq-12467}) \textsc{sq-1,2,4,6,7},
(\subref{fig:sq-12567}) \textsc{sq-1,2,5,6,7},
(\subref{fig:sq-13457}) \textsc{sq-1,3,4,5,7},
(\subref{fig:sq-13467}) \textsc{sq-1,3,4,6,7},
(\subref{fig:sq-13567}) \textsc{sq-1,3,5,6,7},
(\subref{fig:sq-14567}) \textsc{sq-1,4,5,6,7},
(\subref{fig:sq-23457}) \textsc{sq-2,3,4,5,7},
(\subref{fig:sq-23467}) \textsc{sq-2,3,4,6,7},
(\subref{fig:sq-23567}) \textsc{sq-2,3,5,6,7},
(\subref{fig:sq-24567}) \textsc{sq-2,4,5,6,7},
(\subref{fig:sq-34567}) \textsc{sq-3,4,5,6,7},
(\subref{fig:sq-123457}) \textsc{sq-1,2,3,4,5,7},
(\subref{fig:sq-123467}) \textsc{sq-1,2,3,4,6,7},
(\subref{fig:sq-123567}) \textsc{sq-1,2,3,5,6,7},
(\subref{fig:sq-124567}) \textsc{sq-1,2,4,5,6,7},
(\subref{fig:sq-134567}) \textsc{sq-1,3,4,5,6,7},
(\subref{fig:sq-234567}) \textsc{sq-2,3,4,5,6,7},
(\subref{fig:sq-1234567}) \textsc{sq-1,2,3,4,5,6,7}}
\end{figure*}

Here, we calculate the percolation thresholds $p_c$ for random site percolation in the square lattice for neighborhoods containing sites from the 7th coordination zone.
There are 64 such neighborhoods, from \textsc{sq-7} to \textsc{sq-1,2,3,4,5,6,7}, and they are presented in \Cref{fig:neighborhoods}.

The paper is organized as follows: in \Cref{sec:methods} we recall the finite-size scaling hypothesis, present basics of the effective Newman--Ziff Monte Carlo algorithm for the percolation problem, define the effective coordination number and give some details regarding technicalities of computations.
In \Cref{sec:results} we show the results of Monte Carlo simulations which allow for the estimation of 64 values of $p_c$ for various neighborhoods together with their geometrical characteristics, such as total and effective coordination numbers. 
Finally, \Cref{sec:discussion} is devoted to a discussion of the obtained results.

\section{\label{sec:methods}Methods}

\subsection{Finite-Size Scaling Hypothesis}

In the vicinity of the critical point $x_c$, many observables $X$ obey the finite-size scaling \cite{Finite-Size_Scaling_Theory_1990,bookDS,Guide_to_Monte_Carlo_Simulations_2009} relation
\begin{equation}
\label{eq:scaling}
X(x;L)=L^{-\varepsilon_1}\cdot\mathcal F\big((x-x_c)L^{\varepsilon_2}\big),
\end{equation}
where $x$ reflects the level of system disorder, $\mathcal F$ is the scaling function (usually analytically unknown), and $\varepsilon_{1,2}$ are universal exponents.
These exponents depend only on the physical dimension of the system $d$.
Knowing $\varepsilon_{1,2}$ and plotting $X(x;L)$ leads to the collapse of many curves (for various $L$) into a single one. 

At the critical point $x=x_c$, the value of
\begin{equation}
\label{eq:X_at_xc}
X(x=x_c;L)\cdot L^{\varepsilon_1}=\mathcal F\big(0\big)
\end{equation}
does not depend on the size of the system $L$, which opens a computationally feasible way of searching for both the critical point $x_c$ and the critical exponents $\varepsilon_{1,2}$.

For the random site percolation problem, the probability 
\begin{equation}
P_{\text{max}}=S_{\text{max}}/L^2,
\end{equation}
that the randomly chosen site belongs to the largest cluster (with size $S_{\text{max}}$), may play the role of the quantity $X$.
Then, the exponents $\varepsilon_{1,2}$ are known analytically, and $\varepsilon_1=5/48$ while $\varepsilon_2=3/4$. 

\subsection{Efficient Newman--Ziff Algorithm}

To calculate the probability $P_{\text{max}}$ mentioned above we use \citeauthor{NewmanZiff2001} algorithm  \cite{NewmanZiff2001}.
The algorithm is fast as it is based on the concept of calculating desired quantities only after adding a single occupied site to the so far existing system.
This allows us to construct the dependence of $X(n)$, where $n$ is the number of occupied sites.

The second part of the algorithm is the transformation of $X(n)$ values from the space of the integer number $n$ of occupied sites to $X(p)$ in the space of the real numbers of probabilities $p$ of sites occupation
\begin{equation}
\label{eq:Xn-to-Xp}
X(p;N)=\sum_{n=0}^N \bar{X}(n;N) \mathcal B(n;N,p),
\end{equation}
where $N=L^2$ stands for the size of the system.
This conversion requires knowing the Bernoulli (binomial) probability distribution 
\begin{equation}
\label{eq:binom}
\mathcal B(n;N,p)=\binom{N}{n}p^n(1-p)^{N-n}
\end{equation}
and Reference~\cite{NewmanZiff2001} provides an efficient way of recursive calculation of the binomial distribution coefficients in \Cref{eq:binom}.

We implement the \citeauthor{NewmanZiff2001} algorithm as a computer program written in C language.
This requires modification of the original \texttt{boundaries()} procedure provided in Reference~\cite{NewmanZiff2001} beyond the nearest neighbors.
In Listing~\ref{lst:code}, available in Appendix~\ref{app:listing}, we show an example of such modification for the \textsc{sq-7} neighborhood presented in \Cref{{fig:sq-7}}.

\subsection{Effective Coordination Number}

The values of $p_c$ usually degenerate strongly with respect to the neighborhood
coordination number $z$. 
Degeneration means that for a given coordination number (number of sites in the neighborhood), many various values of $p_c$ are associated with this coordination number $z$.
This degeneration may be removed (at least partially) when instead of dealing with the coordination number $z$, we use the effective coordination number 
\begin{equation}
\label{eq:zeta} 
\zeta=\sum_i z_i r_i,
\end{equation}
where $z_i$ and $r_i$ are the number of sites and their distance from the central site in the neighborhood in the $i$-th coordination zone \cite{2204.12593,2303.10423}, respectively. 
Very recently, it was shown that for square, honeycomb and triangular lattices
\begin{equation}
\label{eq:pc_vs_zeta}
p_c(\zeta)\propto\zeta^{-w}
\end{equation}
and exponent $w\approx 1/2$ \cite{2310.20668}.

\section{\label{sec:results}Results}

\begin{figure}[htbp]
\begin{subfigure}[b]{\columnwidth}
\caption{\label{fig:Pmaxvsp-sq-7}}
\includegraphics[width=0.95\columnwidth]{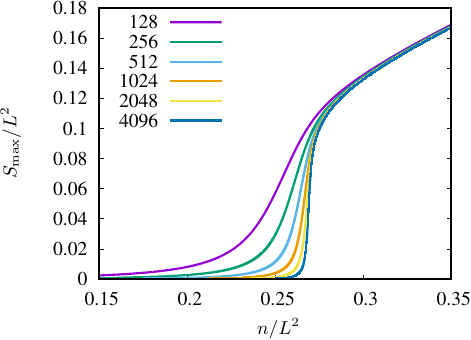}
\end{subfigure}
\begin{subfigure}[b]{\columnwidth}    
\caption{\label{fig:Pmaxvsp-sq-1234567}}
\includegraphics[width=0.95\columnwidth]{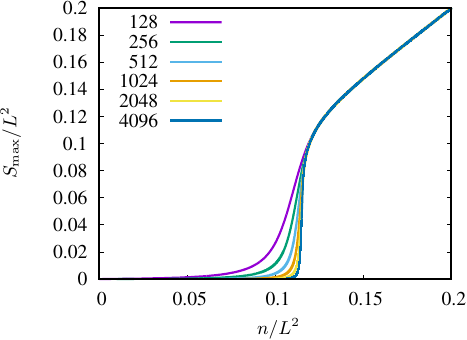}
\end{subfigure}
\caption{\label{fig:Pmaxvsp}Examples of dependencies of $S_{\text{max}}/L^2$ versus $n/L^2$ for various linear system sizes $L=128$, 256, 512, 1024, 2048 and 4096 from top to bottom.
(\subref{fig:Pmaxvsp-sq-7}) \textsc{sq-7},
(\subref{fig:Pmaxvsp-sq-1234567}) \textsc{sq-1,2,3,4,5,6,7}
}
\end{figure}

In \Cref{fig:Pmaxvsp} we present the dependencies of the probabilities $P_{\text{max}}$ of belonging to the largest cluster on sites occupation probability $n/L^2$.
$P_{\text{max}}$ values are defined by geometrical probability as the size of the largest cluster $S_{\text{max}}$ per the system size $L^2$.
The results are averaged over $R=10^5$ realizations of systems that contain $128^2$, $256^2$, $512^2$, $1024^2$, $2048^2$ and $4096^2$ sites.
The examples correspond to the neighborhoods \textsc{sq-7} (\Cref{fig:Pmaxvsp-sq-7}) and \textsc{sq-1,2,3,4,5,6,7} (\Cref{fig:Pmaxvsp-sq-1234567}).

\begin{figure}[htbp]
\begin{subfigure}[b]{\columnwidth}    
\caption{\label{fig:exmpl-PmaxLbetanuvsp-sq-7}}
\includegraphics[width=0.95\columnwidth]{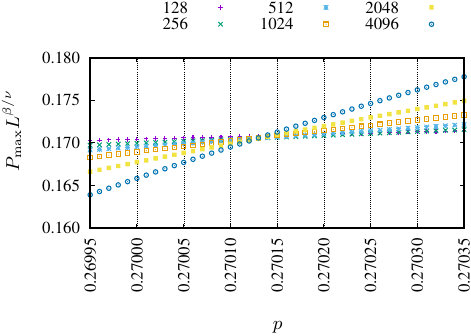}
\end{subfigure}
\begin{subfigure}[b]{\columnwidth}    
\caption{\label{fig:exmpl-PmaxLbetanuvsp-sq-1234567}}
\includegraphics[width=0.95\columnwidth]{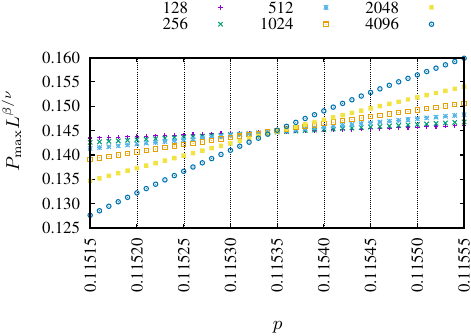}
\end{subfigure}
\caption{\label{fig:exmpl-PmaxLbetanuvsp}Examples of dependencies of $P_{\text{max}}L^{\beta/\nu}$ versus $p$ for various linear system sizes $L=128$, 256, 512, 1024, 2048 and 4096 from top to bottom.
(\subref{fig:exmpl-PmaxLbetanuvsp-sq-7}) \textsc{sq-7}, $p_c(\textsc{sq-7})=0.27013$,
(\subref{fig:exmpl-PmaxLbetanuvsp-sq-1234567}) \textsc{sq-1,2,3,4,5,6,7}, $p_c(\textsc{sq-1,2,3,4,5,6,7})=0.11535$
}
\end{figure}

In \Cref{fig:exmpl-PmaxLbetanuvsp} examples of dependencies of $P_{\text{max}}L^{\beta/\nu}$ versus $p$ are presented for various linear system sizes $L$.
The examples correspond to the neighborhoods \textsc{sq-1} (\Cref{fig:exmpl-PmaxLbetanuvsp-sq-7}) and \textsc{sq-1,2,3,4,5,6,7} (\Cref{fig:exmpl-PmaxLbetanuvsp-sq-1234567}).
These dependencies for 64 neighborhoods containing sites from the 7th coordination zone are presented in Figure~1 in the Supplemental Material \cite{SM}.
The common cross-point for various lattice sizes predicts the percolation threshold $p_c$.

The evaluated percolation thresholds $p_c$, the total ($z$) and the effective ($\zeta$) coordination numbers for various neighborhoods are collected in \Cref{tab:sq-pc}.

\begin{figure}[htbp]
\begin{subfigure}[b]{\columnwidth}
\caption{\label{fig:pc_vs_z}}
\includegraphics[width=0.95\columnwidth]{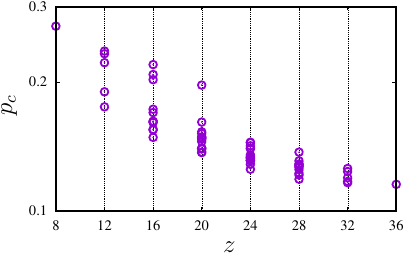}
\end{subfigure}
\begin{subfigure}[b]{\columnwidth}
\caption{\label{fig:pc_vs_zeta}}
\includegraphics[width=0.95\columnwidth]{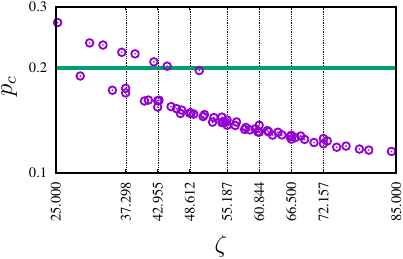}
\end{subfigure}
\begin{subfigure}[b]{\columnwidth}
\caption{\label{fig:pc_vs_zeta_full}}
\includegraphics[width=0.95\columnwidth]{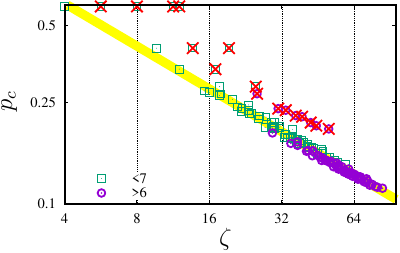}
\end{subfigure}
\caption{\label{fig:pc_vs_z_zeta}Dependencies of $p_c$ on (\subref{fig:pc_vs_z}) the total ($z$) and (\subref{fig:pc_vs_zeta}) effective ($\zeta$) coordination number for neighborhoods presented in \Cref{fig:neighborhoods}.
(\subref{fig:pc_vs_zeta_full}) $p_c(\zeta)$ with additional data also for neighborhoods (marked by {\color{green}$\square$}) containing sites up to the 6th coordination zone (after Reference~\cite{2310.20668} and references therein).
The inflated neighborhoods (marked by {\color{red}$\times$}) are excluded from the fitting procedure.
The least-squares fit gives exponent $w\approx 0.5552(42)$}
\end{figure}

In \Cref{fig:pc_vs_z_zeta} the dependencies of the percolation threshold $p_c$ on the total ($z$, \Cref{fig:pc_vs_z}) and effective ($\zeta$, \Cref{fig:pc_vs_zeta}) are presented.
\Cref{fig:pc_vs_zeta_full} shows $p_c(\zeta)$ with additional data also for neighborhoods containing sites up to the 6th coordination zone \cite{2310.20668}.
The inflated neighborhoods (marked by $\times$) are excluded from the fitting procedure.
The data are accompanied by the least squares method fit to the power law \eqref{eq:pc_vs_zeta}.
The estimated value of the exponent $w\approx 0.5552(42)$.

\begingroup
\squeezetable
\begin{table}[!h]
\caption{\label{tab:sq-pc}Percolation thresholds $p_c$ for a square lattice with complex neighborhoods (and their characteristics $z$, $\zeta$) containing sites from the 7th coordination zone presented in \Cref{fig:neighborhoods}}
\begin{ruledtabular}
\begin{tabular}{lddd}
	lattice & \text{$z$} & \text{$\zeta$} & \text{$p_c$}\\
	lattice & \text{$z$} & \text{$\zeta$} & \text{$p_c$}\\
\hline 
\textsc{sq-1,2,3,4,5,6,7} & 36 & 84.157 & 0.11535\\
\hline 
\textsc{sq-2,3,4,5,6,7}   & 32 & 80.157 & 0.11636\\
\textsc{sq-1,3,4,5,6,7}   & 32 & 78.500 & 0.11719\\
\textsc{sq-1,2,4,5,6,7}   & 32 & 76.157 & 0.11949\\
\textsc{sq-1,2,3,5,6,7}   & 28 & 66.269 & 0.12725\\
\textsc{sq-1,2,3,4,6,7}   & 32 & 72.844 & 0.12358\\
\textsc{sq-1,2,3,4,5,7}   & 32 & 72.157 & 0.12562\\
\hline 
\textsc{sq-3,4,5,6,7}     & 28 & 74.500 & 0.11859\\
\textsc{sq-2,4,5,6,7}     & 28 & 72.157 & 0.12132\\
\textsc{sq-2,3,5,6,7}     & 24 & 62.269 & 0.13241\\
\textsc{sq-2,3,4,6,7}     & 28 & 68.844 & 0.12488\\
\textsc{sq-2,3,4,5,7}     & 28 & 68.157 & 0.12770\\
\textsc{sq-1,4,5,6,7}     & 28 & 70.500 & 0.12236\\
\textsc{sq-1,3,5,6,7}     & 24 & 60.612 & 0.13112\\
\textsc{sq-1,3,4,6,7}     & 28 & 67.187 & 0.12644\\
\textsc{sq-1,3,4,5,7}     & 28 & 66.500 & 0.12830\\
\textsc{sq-1,2,5,6,7}     & 24 & 58.269 & 0.13339\\
\textsc{sq-1,2,4,6,7}     & 28 & 64.844 & 0.12864\\
\textsc{sq-1,2,4,5,7}     & 28 & 64.157 & 0.13089\\
\textsc{sq-1,2,3,6,7}     & 24 & 54.955 & 0.13973\\
\textsc{sq-1,2,3,5,7}     & 24 & 54.269 & 0.14470\\
\textsc{sq-1,2,3,4,7}     & 28 & 60.844 & 0.13718\\
\hline 
\textsc{sq-4,5,6,7}       & 24 & 66.500 & 0.12513\\
\textsc{sq-3,5,6,7}       & 20 & 56.612 & 0.13689\\
\textsc{sq-3,4,6,7}       & 24 & 63.187 & 0.12848\\
\textsc{sq-3,4,5,7}       & 24 & 62.500 & 0.13121\\
\textsc{sq-2,5,6,7}       & 20 & 54.269 & 0.13959\\
\textsc{sq-2,4,6,7}       & 24 & 60.844 & 0.13104\\
\textsc{sq-2,4,5,7}       & 24 & 60.157 & 0.13386\\
\textsc{sq-2,3,6,7}       & 20 & 50.955 & 0.14523\\
\textsc{sq-2,3,5,7}       & 20 & 50.269 & 0.19672\\
\textsc{sq-2,3,4,7}       & 24 & 56.844 & 0.14015\\
\textsc{sq-1,5,6,7}       & 20 & 52.612 & 0.13998\\
\textsc{sq-1,4,6,7}       & 24 & 59.187 & 0.13298\\
\textsc{sq-1,4,5,7}       & 24 & 58.500 & 0.13515\\
\textsc{sq-1,3,6,7}       & 20 & 49.298 & 0.14760\\
\textsc{sq-1,3,5,7}       & 20 & 48.612 & 0.14876\\
\textsc{sq-1,3,4,7}       & 24 & 55.187 & 0.14187\\
\textsc{sq-1,2,6,7}       & 20 & 46.955 & 0.14814\\
\textsc{sq-1,2,5,7}       & 20 & 46.269 & 0.15298\\
\textsc{sq-1,2,4,7}       & 24 & 52.844 & 0.14423\\
\textsc{sq-1,2,3,7}       & 20 & 42.955 & 0.16125\\
\hline 
\textsc{sq-5,6,7}         & 16 & 48.612 & 0.14848\\
\textsc{sq-4,6,7}         & 20 & 55.187 & 0.13708\\
\textsc{sq-4,5,7}         & 20 & 54.500 & 0.14008\\
\textsc{sq-3,6,7}         & 16 & 45.298 & 0.15503\\
\textsc{sq-3,5,7}         & 16 & 44.612 & 0.20250\\
\textsc{sq-3,4,7}         & 20 & 51.187 & 0.14709\\
\textsc{sq-2,6,7}         & 16 & 42.955 & 0.15461\\
\textsc{sq-2,5,7}         & 16 & 42.269 & 0.20831\\
\textsc{sq-2,4,7}         & 20 & 48.844 & 0.14868\\
\textsc{sq-2,3,7}         & 16 & 38.955 & 0.21963\\
\textsc{sq-1,6,7}         & 16 & 41.298 & 0.16193\\
\textsc{sq-1,5,7}         & 16 & 40.612 & 0.16095\\
\textsc{sq-1,4,7}         & 20 & 47.187 & 0.15157\\
\textsc{sq-1,3,7}         & 16 & 37.298 & 0.16973\\
\textsc{sq-1,2,7}         & 16 & 34.955 & 0.17278\\
\hline 
\textsc{sq-6,7}           & 12 & 37.298 & 0.17497\\
\textsc{sq-5,7}           & 12 & 36.612 & 0.22190\\
\textsc{sq-4,7}           & 16 & 43.187 & 0.16171\\
\textsc{sq-3,7}           & 12 & 33.298 & 0.23288\\
\textsc{sq-2,7}           & 12 & 30.955 & 0.23619\\
\textsc{sq-1,7}           & 12 & 29.298 & 0.18976\\
\hline 
\textsc{sq-7}             &  8 & 25.298 & 0.27013\\
\end{tabular}
\end{ruledtabular}
\end{table}

\section{\label{sec:discussion}Discussion}

In this paper, the estimation of $p_c$ for random site percolation on the square lattice for neighborhoods containing sites from the 7th coordination zone is presented.
Monte Carlo simulation with $R=10^5$ system realizations for the system sizes from $128^2$ to $4096^2$ sites allowed us to estimate the percolation thresholds to the $10^{-5}$ accuracy.

The obtained percolation thresholds range from $0.27013$ (for the neighborhood that contains solely sites from the 7th coordination zone) to $0.11535$ (for the neighborhood that contains all sites from the 1st to the 7th coordination zone).
The latter agrees perfectly (with the accuracy obtained here, that is $10^{-5}$) with the earlier results of the extensive Monte Carlo simulation (with $L=16384$ and $R>3\cdot 10^8$ independent samples produced for each lattice) where $p_c(\textsc{sq-1,2,3,4,5,6,7})\approx 0.1153481(9)$ \cite{PhysRevE.103.022126}.

Capturing the intersection of the rescaled probabilities of belonging to the largest cluster $L^{\beta/\nu}\cdot P_{\text{max}}(p)$ numerically for finite systems is quite challenging as the curves for various $L$ rather seldom intersect exactly at one point (as predicted theoretically by \Cref{eq:X_at_xc}). 
Also in our case, in \Cref{fig:exmpl-PmaxLbetanuvsp}, we do not have the intersection of the $L^{\beta/\nu}\cdot P_{\text{max}}(p)$ curves for different $L$ occurring at one point, but rather we can easily identify the value of $p^*$ for which the mutual squared differences 
\[
\delta^2(p)=\sum_{i,j} \left[ L_i^{\beta/\nu}\cdot P_\text{max}(p;L_i)-L_j^{\beta/\nu}\cdot P_\text{max}(p;L_j)\right] ^2
\] 
between the $L^{\beta/\nu}\cdot P_{\text{max}}(p)$ values---for all studied $L$---is the smallest.
This value of $p^*$ for which the mutual squared differences $\delta^2(p)$ reaches its minimum estimates the percolation threshold $p_c\approx p^*$ \cite{Bastas2014}.
The convolution \eqref{eq:Xn-to-Xp} can be performed for any arbitrary value of $p$, but the secret of achieving such a small $\delta^2$ value that one has the impression it tends to zero lies in the reasonable assumption of the separation $\Delta p$ with which we scan $p$ values. 
In \Cref{fig:errors} we show examples of $L^{\beta/\nu}\cdot P_{\text{max}}(p)$ for $\Delta p=10^{-4}$, $10^{-5}$ and $10^{-6}$. 
As can be seen, these dependencies allow us to easily indicate the ``intersection point'' for $\Delta p=10^{-4}$ (\Cref{fig:errors_dp-4}) and $10^{-5}$ (\Cref{fig:errors_dp-5}).
That is, just by inspection, even without calculating $\delta^2$, one can see a higher dispersion of the values $L^{\beta/\nu}\cdot P_{\text{max}}(p)$ for the considered values of $L$ and thus larger $\delta^2(p)$ at points $p=(p^*-\Delta p)$ and $p=(p^*+\Delta p)$ than at $p=p^*$.
This means that the true value of $p_c$ is somewhere inside the interval $(p^*-\Delta p,p^*+\Delta p)$.

On the other hand, for $\Delta p=10^{-6}$ (see \Cref{fig:errors_dp-6}) we cannot easily identify the ``intersection point''.
The reason for this lies in too weak statistics, i.e., too small number $R$ of the simulation repetition. 
The precision of determining the value of $P_\text{max}(n)$---i.e. $\bar X(n;N)$ in \Cref{eq:Xn-to-Xp}--is $1/R$ and affects the selection of $\Delta p$ value, which allows observing a clear $\delta^2(p)$ minimum for $p=p^*$, with a simultaneous clear spread of points $L^{\beta/\nu}\cdot P_\text{max}(p)$ for various $L$ at $p=p^*-\Delta p$ and $p=p^*+\Delta p$. 
Hence, we consider $\Delta p=10^{-5}$ as the uncertainty of the determined $p_c$. 
Moreover, plotting $L^{\beta/\nu}\cdot P_\text{max}(p)$ for finite $L$ according to \Cref{eq:scaling} allows us to eliminate, at least partially, the effect of finite sizes on the accuracy of determining $p_c$. 

In \Cref{fig:pc_vs_zeta} we clearly observe two series of data, roughly for $p_c$ below and above 0.2.
The latter correspond to \textsc{sq-7}, \textsc{sq-2,7}, \textsc{sq-3,7}, \textsc{sq-5,7}, \textsc{sq-2,3,7}, \textsc{sq-2,5,7}, \textsc{sq-3,5,7} and \textsc{sq-2,3,5,7}.
These neighborhoods are the so-called inflated neighborhoods, which means that they have partners with lower indexed neighborhoods with the same $p_c$ and $z$. 
These partners are presented in \Cref{fig:inflated}.

Similarly to neighborhoods with smaller ranges, the power-law dependence of the percolation threshold on the effective coordination number with exponent $w$ close to $-1/2$ is observed.
The power law is obtained after excluding $p_c$ of the inflated neighborhoods (marked by $\times$ in \Cref{fig:pc_vs_zeta_full}) from the fitting procedure.

Introducing the effective coordination number $\zeta$ partially eliminates the degeneration (see \Cref{fig:pc_vs_zeta}) strongly observed in the dependence on $p_c(z)$ (see \Cref{fig:pc_vs_z}).
For $p_c(\zeta)$, we still observe this degeneration in several cases, such as
\begin{itemize}
    \item $\zeta(\textsc{sq-1,3,7})=\zeta(\textsc{sq-6,7})$,  
    \item $\zeta(\textsc{sq-1,2,3,7})=\zeta(\textsc{sq-2,6,7})$, 
    \item $\zeta(\textsc{sq-1,3,5,7})=\zeta(\textsc{sq-5,6,7})$, 
    \item $\zeta(\textsc{sq-1,3,4,7})=\zeta(\textsc{sq-4,6,7})$, 
    \item $\zeta(\textsc{sq-1,2,3,4,7})=\zeta(\textsc{sq-2,4,6,7})$,
    \item $\zeta(\textsc{sq-1,3,4,5,7})=\zeta(\textsc{sq-4,5,6,7})$,
    \item $\zeta(\textsc{sq-1,2,3,4,5,7})=\zeta(\textsc{sq-2,4,5,6,7})$.
\end{itemize}

This degeneration in distinguishing neighborhoods based only on the scalar variable can be resolved after normalization of $\zeta$ to the total number of sites in the neighborhood $z$.
The fraction $\zeta/z$ is nothing else but the mean distance
\begin{equation}
\bar r = \frac{1}{z} \sum_i z_ir_i
\end{equation}
of sites in the neighborhood to the central site.
We can call it the mean radius of the neighborhood.

\begin{figure}[htbp]
\begin{subfigure}[b]{\columnwidth}
\caption{\label{fig:errors_dp-4}}
\includegraphics[width=0.95\columnwidth]{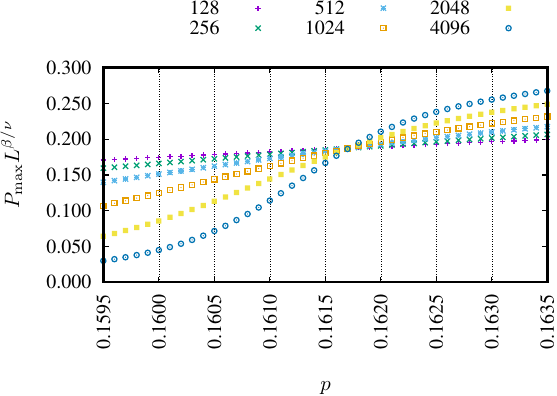}
\end{subfigure}
\hfill 
\begin{subfigure}[b]{\columnwidth}
\caption{\label{fig:errors_dp-5}}

\includegraphics[width=0.95\columnwidth]{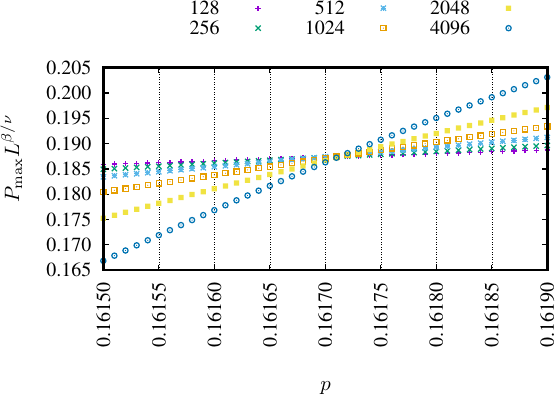}
\end{subfigure}
\hfill 
\begin{subfigure}[b]{\columnwidth}
\caption{\label{fig:errors_dp-6}}

\includegraphics[width=0.95\columnwidth]{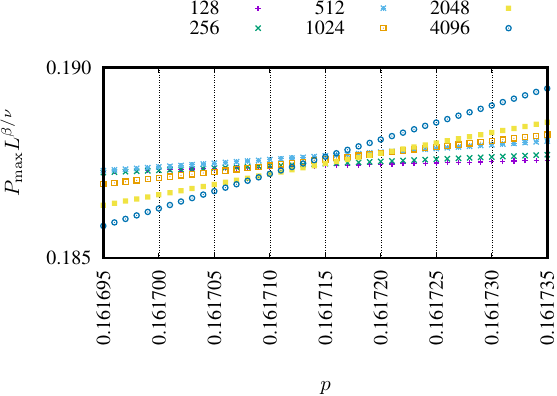}
\end{subfigure}
\caption{\label{fig:errors}Dependencies $L^{\beta/\nu}\cdot P_\text{max}(p)$ for various values of the separation $\Delta p$:
(\subref{fig:errors_dp-4}) $10^{-4}$,
(\subref{fig:errors_dp-5}) $10^{-5}$,
(\subref{fig:errors_dp-6}) $10^{-6}$.
With assumed number of repetition of simulations ($R=10^5$) value $\Delta p=10^{-6}$ is insufficient to see clear ``single crossing-point'' and to identify $p=p^*$}
\end{figure}

\begin{figure}[htbp]
\begin{subfigure}[b]{0.107\textwidth}
\caption{\label{fig:inf-sq-7}}
\includegraphics[width=\textwidth]{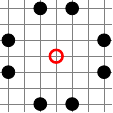}
\end{subfigure}
\hspace{1cm} 
\begin{subfigure}[b]{0.107\textwidth}
\caption{\label{fig:sq-4}}
\includegraphics[width=\textwidth]{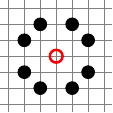}
\end{subfigure}\\
\begin{subfigure}[b]{0.107\textwidth}
\caption{\label{fig:inf-sq-27}}
\includegraphics[width=\textwidth]{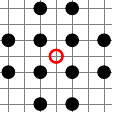}
\end{subfigure}
\hspace{1cm} 
\begin{subfigure}[b]{0.107\textwidth}
\caption{\label{fig:sq-14}}
\includegraphics[width=\textwidth]{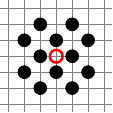}
\end{subfigure}\\
\begin{subfigure}[b]{0.107\textwidth}
\caption{\label{fig:inf-sq-37}}
\includegraphics[width=\textwidth]{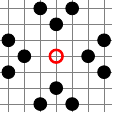}
\end{subfigure}
\hspace{1cm} 
\begin{subfigure}[b]{0.107\textwidth}
\caption{\label{fig:sq-24}}
\includegraphics[width=\textwidth]{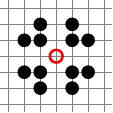}
\end{subfigure}\\
\begin{subfigure}[b]{0.107\textwidth}
\caption{\label{fig:inf-sq-57}}
\includegraphics[width=\textwidth]{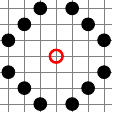}
\end{subfigure}
\hspace{1cm} 
\begin{subfigure}[b]{0.107\textwidth}
\caption{\label{fig:sq-34}}
\includegraphics[width=\textwidth]{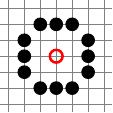} 
\end{subfigure}\\
\begin{subfigure}[b]{0.107\textwidth}
\caption{\label{fig:inf-sq-237}}
\includegraphics[width=\textwidth]{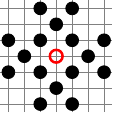}
\end{subfigure}
\hspace{1cm} 
\begin{subfigure}[b]{0.107\textwidth}
\caption{\label{fig:sq-124}}
\includegraphics[width=\textwidth]{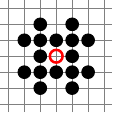}
\end{subfigure}\\
\begin{subfigure}[b]{0.107\textwidth}
\caption{\label{fig:inf-sq-257}}
\includegraphics[width=\textwidth]{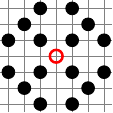}
\end{subfigure}
\hspace{1cm} 
\begin{subfigure}[b]{0.107\textwidth}
\caption{\label{fig:sq-134}}
\includegraphics[width=\textwidth]{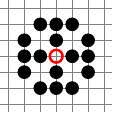}
\end{subfigure}\\ 
\begin{subfigure}[b]{0.107\textwidth}
\caption{\label{fig:inf-sq-357}}
\includegraphics[width=\textwidth]{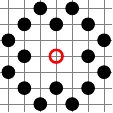}
\end{subfigure}
\hspace{1cm} 
\begin{subfigure}[b]{0.107\textwidth}
\caption{\label{fig:sq-234}}
\includegraphics[width=\textwidth]{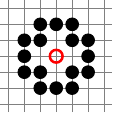}
\end{subfigure}\\
\begin{subfigure}[b]{0.107\textwidth}
\caption{\label{fig:inf-sq-2357}}
\includegraphics[width=\textwidth]{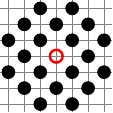}
\end{subfigure}
\hspace{1cm} 
\begin{subfigure}[b]{0.107\textwidth}
\caption{\label{fig:sq-1234}}
\includegraphics[width=\textwidth]{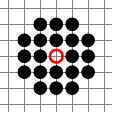}
\end{subfigure}
\caption{\label{fig:inflated}Inflated neighborhoods 
(\subref{fig:inf-sq-7}) \textsc{sq-7},
(\subref{fig:inf-sq-27}) \textsc{sq-2,7},
(\subref{fig:inf-sq-37}) \textsc{sq-3,7},
(\subref{fig:inf-sq-57}) \textsc{sq-5,7},
(\subref{fig:inf-sq-237}) \textsc{sq-2,3,7},
(\subref{fig:inf-sq-257}) \textsc{sq-2,5,7},
(\subref{fig:inf-sq-357}) \textsc{sq-3,5,7},
(\subref{fig:inf-sq-2357}) \textsc{sq-2,3,5,7}
and their lower indexed partners
(\subref{fig:sq-4}) \textsc{sq-4},
(\subref{fig:sq-14}) \textsc{sq-1,4},
(\subref{fig:sq-24}) \textsc{sq-2,4},
(\subref{fig:sq-34}) \textsc{sq-3,4},
(\subref{fig:sq-124}) \textsc{sq-1,2,4},
(\subref{fig:sq-134}) \textsc{sq-1,3,4},
(\subref{fig:sq-234}) \textsc{sq-2,3,4},
(\subref{fig:sq-1234}) \textsc{sq-1,2,3,4}}
\end{figure}

In \Cref{fig:pc_vs_r} we show the dependence $p_c$ on the mean radius $\bar r$ for the 131 neighborhoods that contain sites in the range smaller than or equal to $\bar r(\textsc{sq-7})=\sqrt{10}$.
In this plot, only three neighborhoods have identical mean radius $\bar r(\textsc{sq-3}) = \bar r(\textsc{sq-1,6}) = \bar r(\textsc{sq-1,3,6}) = 2$ (these three neighborhoods are marked by $\times$).
Furthermore, we can empirically determine the lower limit of the percolation threshold for complex neighborhoods as
\begin{equation}
\label{eq:limit}
p_c(\textsc{sq}) \ge \frac{p_c(\textsc{sq-1})}{\bar r^2}.
\end{equation}
This limit also holds for extended neighborhoods with sites beyond the 7th coordination zone, for example \textsc{sq-1,2,3,4,5,6,7,8}, \textsc{sq-1,2,3,4,5,6,7,8,9} and \textsc{sq-1,2,3,4,5,6,7,8,9,10} ($p_c$ values for these neighborhoods are taken after Reference~\cite{PhysRevE.105.024105}).
The results for extended neighborhoods (which are both complex and compact, marked by $+$ in \Cref{fig:pc_vs_r}) touch the boundary line of inequality \eqref{eq:limit}.

\begin{figure}[htbp]
\includegraphics[width=0.95\columnwidth]{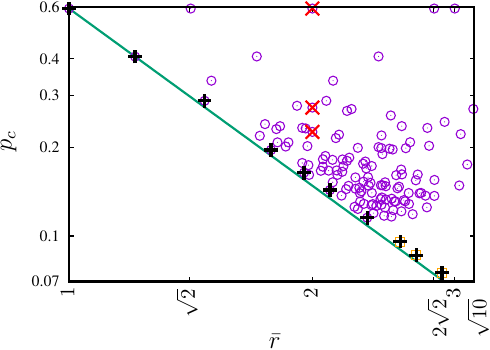}
\caption{\label{fig:pc_vs_r}Dependence of percolation threshold $p_c$ on mean radius $\bar r$ of the neighborhoods. The mean radius $\bar r$ uniquely identifies neighborhoods except of three neighborhoods \textsc{sq-3}, \textsc{sq-1,6} and \textsc{sq-1,3,6} all with mean radius equal to two. These three neighborhoods are marked by crosses ($\times$).
The open squares ($\square$) correspond to \textsc{sq-1,2,3,4,5,6,7,8}, \textsc{sq-1,2,3,4,5,6,7,8,9}, \textsc{sq-1,2,3,4,5,6,7,8,9,10} neighborhoods (after Reference~\cite{PhysRevE.105.024105}).
Pluses ($+$) show $p_c$ associated with compact neighborhoods}
\end{figure}

Also in Reference~\cite{PhysRevE.105.024105}---from which we took values of $p_c$ for \textsc{sq-1,2,3,4,5,6,7,8}, \textsc{sq-1,2,3,4,5,6,7,8,9}, \textsc{sq-1,2,3,4,5,6,7,8,9,10} neighborhoods---\citeauthor{PhysRevE.105.024105} studied, among other things, also the percolation thresholds for regular lattices with compact extended-range neighborhoods in two dimensions.
For all Archimedean lattices and up to the 10th nearest neighbors, they show the dependence $z$ versus $1/p_c$ (Figure~7 in Reference~\cite{PhysRevE.105.024105}) and $-1/\ln(1-p_c)$ (Figure~8 in Reference~\cite{PhysRevE.105.024105}). 
For new variables $y=z$ and $x=1/p_c$ or $x=-1/\ln(1-p_c)$ in both cases the slope of the straight line close to the experimental points is $y=4.521x$ and the value of $4.521=4\eta_c$ comes from the critical filling factor of circular neighborhoods in two dimensions (i.e. for continuous percolation of discs, where $\eta_c=1.12808737(6)$ \cite{PhysRevE.86.061109}). The experimental data for $z$ vs. $1/p_c$ lie below this straight line as compact neighborhoods become solid discs in the limit of $z\to\infty$.
In \Cref{fig:pc_vs_z_r2} the reciprocal of $p_c$ both from our work (against $r^2$) and the continuous percolation limit (from Reference~\cite{PhysRevE.105.024105}, against $z$) are presented.
As we can see, for compact neighborhoods (at least for the square lattice and site percolation problem), we can confine percolation thresholds $p_c$ between two curves.

To conclude, we calculated 64 percolation thresholds for neighborhoods containing sites from the 7th coordination zone, of which 63 are evaluated for the first time.
The obtained values of $p_c$ follow the early prediction of $p_c(\zeta)$, which is given by the power law $p_c \propto \zeta^{-w}$ with the exponent $w$ close to $1/2$.
Investigating the degeneration of $p_c$ versus $\zeta$ allowed us to determine the lower limit $p_c$ as dependent on the inverse square of the mean distance $\bar r$ of sites in the neighborhoods.
The latter touches the boundary line for the extended (compact) neighborhoods.
These results enrich earlier studies of site percolation for compact neighborhoods \cite{PhysRevE.105.024105} where $p_c$ values were restricted by the limitation predicted by $1/p_c>z/(4\eta_c)$, where $\eta_c$ is the critical filling factor for the continuous percolation of discs.
Finally, we also recalculated $p_c(\textsc{sq-2,4})=0.23288$, which means that its value provided in Reference~\cite{Majewski2007} as $p_c(\textsc{sq-2,4})=0.225$ was clearly underestimated.

\begin{figure}[htbp]
\includegraphics[width=0.95\columnwidth]{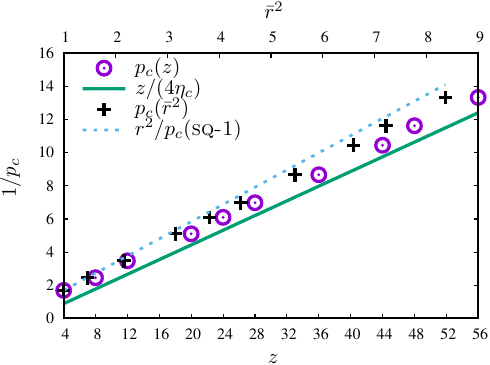}
\caption{\label{fig:pc_vs_z_r2}Dependence of reciprocal percolation threshold $1/p_c$ on $\bar r^2$ and $z$ of the neighborhoods.
Data for \textsc{sq-1,2,3,4,5,6,7,8}, \textsc{sq-1,2,3,4,5,6,7,8,9}, \textsc{sq-1,2,3,4,5,6,7,8,9,10} neighborhoods are taken from Reference~\cite{PhysRevE.105.024105} together with the continuous percolation limit of the discs ($4\eta_c$)}
\end{figure}

Further studies may concentrate on the estimation of the percolation thresholds $p_c$ for triangular or honeycomb lattices with complex neighborhoods containing sites from the 7th coordination zone or validation of \Cref{eq:limit} for other lattices.

\acknowledgments{Authors gratefully acknowledge Poland’s high-performance computing infrastructure \href{https://ror.org/01m3qaz74}{PLGrid} (HPC Centers: ACK Cyfronet AGH) for providing computer facilities and support within computational grant no.~PLG/2023/016295.}

\appendix
\section[\appendixname~\thesection]{\label{app:listing}}

In Listing~\ref{lst:code} the \texttt{boundaries()} procedure to be replaced in the original program published in Reference~\cite{NewmanZiff2001} is presented.
The example corresponds to the \textsc{sq-7} neighborhood.

\begin{lstlisting}[language=c,label=lst:code,caption=\texttt{boundaries()} procedure for \textsc{sq-7} neighborhood to be inserted in Newman--Ziff algorithm code published in Reference~\cite{NewmanZiff2001}]
// SQ-7
#define L 9       /* Linear dimension */
#define Z 8
#define N (L*L)

int nn[N][Z];     /* Nearest neighbors */

void boundaries() {
  int i,j;
  for(i=0; i<N; i++) {
    // 7nn core:
    nn[i][0] = (N+i   +L+3)%N;
    nn[i][1] = (N+i   +L-3)%N;  
    nn[i][2] = (N+i   -L+3)%N;
    nn[i][3] = (N+i   -L-3)%N; 
    nn[i][4] = (N+i +3*L+1)%N; 
    nn[i][5] = (N+i +3*L-1)%N;
    nn[i][6] = (N+i -3*L+1)%N; 
    nn[i][7] = (N+i -3*L-1)%N; 
    // 7nn left border:
    if(i%L==0) {
       nn[i][1] = (N+L+i   +L-3)%N;
       nn[i][3] = (N+L+i   -L-3)%N;
       nn[i][5] = (N+L+i +3*L-1)%N;
       nn[i][7] = (N+L+i -3*L-1)%N; }
    if(i%L==1 || i%L==2) {
       nn[i][1] = (N+L+i   +L-3)%N;
       nn[i][3] = (N+L+i   -L-3)%N; }
    // 7nn right border:
    if((i+3)%L==0 || (i+2)%L==0) {
       nn[i][0] = (N-L+i   +L+3)%N;
       nn[i][2] = (N-L+i   -L+3)%N; }
    if((i+1)%L==0) {
       nn[i][0] = (N-L+i   +L+3)%N;
       nn[i][2] = (N-L+i   -L+3)%N;
       nn[i][4] = (N-L+i +3*L+1)%N;
       nn[i][6] = (N-L+i -3*L+1)%N; }
  }
}
\end{lstlisting}

%
\end{document}